\documentclass[letterpaper, 10 pt, conference]{ieeeconf}  

\IEEEoverridecommandlockouts                              

\usepackage{amsfonts}

\title{\LARGE \bf
Sensor Importance towards Observability Degree via Shapley Values
}

\author{Vishal Cholapadi Ravindra$^{*}$
\thanks{$^{*}$ Vishal Cholapadi Ravindra is with Intuit Credit Karma, Bengaluru, India
        {\tt\small vishal.ravindra@creditkarma.com} \newline
        © 2025 IEEE. Personal use of this material is permitted. Permission from IEEE must be obtained for all other uses, in any current or future media, including reprinting/republishing this material for advertising or promotional purposes, creating new collective works, for resale or redistribution to servers or lists, or reuse of any copyrighted component of this work in other works. This paper has been accepted for publication in the Proceedings of the Eleventh Indian Control Conference (ICC 2025), held in Bengaluru, India on December 18-20, 2025.}%
}

\begin{document}

\maketitle
\thispagestyle{empty}
\pagestyle{empty}

\begin{abstract}

Sensor selection is an often under-appreciated aspect of state estimator or Kalman filter design. The basic minimum requirement for the choice of a sensor set while designing Kalman filters is that all states are observable. In addition, the sensors should be chosen with a view towards estimating the states with a desired accuracy. Often observability is treated as true/false check during filter design. Beyond observability --- the observability degree --- which measures \emph{how observable} the states are, has been used as the metric of choice to for sensor selection or placement applications. The higher the degree of observability, the better the possibility of designing Kalman filters that achieve the desired state estimation accuracy and consistency requirements. When a wide variety of sensors are available, sometimes with cost and physical constraints involved, sensor selection plays a crucial role in filter design.  In such situations it is important to know the \emph{expected} contribution of each sensor towards observability degree. Shapley values, developed in cooperative game theory for fair allocation of the payout of a multi-player game to individual players, are widely used in machine learning to assess feature importance. This paper shows that Shapley values can indeed be leveraged to quantify the \emph{expected marginal} contribution of each sensor in any given sensor set towards the observability degree. This quantification of the fair contribution of each sensor towards the observability degree can be leveraged by filter designers for sensor selection, placement and filter (state estimator) design.

\end{abstract}

\section{INTRODUCTION}
A deterministic state space system is said to be \emph{fully observable} if its initial state can be fully and uniquely recovered from a finite number of observations of the output and knowledge of the input. As a deterministic system is not affected by noise, knowledge of the initial state is equivalent to the knowledge of the state at any instance of time \cite{ybs-purple-book}. Generalizing this further, suppose the state dynamics and the observations of the output are modeled as driven by zero mean additive white Gaussian noise, it is well known that the Kalman filter (\cite{kalman60}) provides the optimal minimum mean square estimate of the unknown state \cite{ybs-purple-book}. When the state is modeled as stochastic, observability is not just a sufficient but also a necessary condition to design a filter that can converge over time. 

The two most common ways to check if a system is observable is via the observability matrix and the observability Gramian. The system is observable if the observability matrix is full column rank \cite{ybs-purple-book} or if the observability Gramian is not singular \cite{kailath80}. These two approaches however tell us only if or not if the system is observable, and do not quantify the \emph{degree of observability}. Degree of observability\footnote{Observability degree and degree of observability are used interchangeably in this paper.} is defined as a measure of \emph{how observable an observable system is}. For fully observable systems, the degree of observability is an important criteria for applications such as sensor selection/  placement \cite{vita2014, summers2015}, path planning \cite{rafiesakhaei2017} and Kalman filter design \cite{ruggaber2021}.  A combination of sensors is more desirable from the point of view of filter design if it significantly improves the degree of observability over alternate combinations. Several metrics based on the observability Gramian matrix, such as trace, rank, log determinant, condition number, etc. have been proposed and evaluated in \cite{krener09}--\cite{staiger23}. The essence being the evaluation of \emph{how close} the observability Gramian matrix is to being singular. 

Problems like sensor selection, sensor placement and virtual sensing (\cite{peterson08, summers2015, wenzel07}) focus on the selection of the ``optimal'' set of sensors that are able to achieve a desired degree of observability. Implications of the modularity properties of observability Gramian based metrics are discussed in \cite{summers2015},\footnote{The author acknowledges the anonymous reviewer who pointed them towards this reference.} where it is shown that simple greedy approaches can be used to solve NP-hard optimal sensor placement problems if the metrics are submodular in nature. A Kalman filter measurement update equation based approach was proposed in \cite{ruggaber2021}, for attributing the impact of each sensor on the estimation of individual states. However, this approach is predicated upon the running of a Kalman filter using a given sensor set.

The aim of the present paper is not the solution of the ``optimal'' sensor selection or placement problem, but to introduce a way for fair quantification of the contribution of each sensor in an overall coalition or set of sensors being considered for filter design, towards selected observability degree measures. Sometimes it is important to \emph{exactly} know what is the individual contribution of each sensor, especially in cases such as virtual sensing where the goal is to replace sensors of higher cost with virtual sensors or state estimates. Furthermore, the method proposed is not conditional on filter design like in \cite{ruggaber2021}, but rather can be performed before it.

The present paper proposes to leverage a concept first proposed in cooperative game theory for \emph{fair attribution} of the payoff of a multi-player game to each individual player \cite{shapley53}. The Shapley value of a player\footnote{A player is analogous to a sensor, while a multi-player game is analogous to the sensor set being considered.} is the marginal contribution of a player to the overall payout which is characterized by a value function. Shapley values are widely used in machine learning \cite{molnar2025} to understand feature importance in order to explain individual predictions from a classification or a regression model. The contribution of this paper is to show that Shapley values can \emph{also} be applied to observability degree measures (based on the observability Gramian) in order to arrive at fair and exact contributions of individual sensors.  Shapley values can be derived based on any observability degree metric, irrespective of their modularity properties (\cite{summers2015}) as long as certain smoothness conditions are met \cite{molnar2025}. The appeal of this approach is two fold: (a) avoiding the need to run a Kalman filter \emph{a priori} and (b) \emph{interpretability} -- as it is known that Shapley values are a fair attribution metric \cite{molnar2025, shapley53}. As the present paper is intended to serve as a concept introduction paper, the scope is limited to linear time invariant (LTI) systems. Using two simple simulation scenarios, the ability of Shapley values to learn interaction effects between individual sensors that enhance the observability degree is highlighted.

Section \ref{sec:state-space-obs} provides a concise mathematical background for LTI state space systems, observability and measures of characterizing observability. Section \ref{sec:deg-observability} introduces the concept of degree of observability and describes two commonly used measures for it. Section \ref{sec:shapley-values} introduces Shapley values from cooperative game theory and describes the four axioms that are satisfied by them in order for them to be considered fair allocations of payouts for each player in a multi-player game. It has to be noted that the complexity of exact Shapley value computation grows exponentially with the number of sensors in the full coalition being considered. However, leveraging of fast approximation techniques are left out of scope of this paper and can instead be found in \cite{molnar2025}. Section \ref{sec:shapley-observability} describes the key contribution of this paper, which is to apply Shapley values to quantify the individual contribution of each sensor towards observability degree. Section \ref{sec:sim-results} describes two simulation scenarios and results that demonstrate the idea described in the present paper, followed by Conclusions.

\section{State space system and observability}
\label{sec:state-space-obs}
Consider the following (deterministic) discrete time linear time invariant (LTI) system 
\begin{eqnarray}
	\textbf{x}_{k+1} &=& A\textbf{x}_k \nonumber \\
	\textbf{y}_{k} &=& C\textbf{x}_{k}  \label{eq:state-space-system}
\end{eqnarray}
where $\textbf{x} \in \mathbb{R}^n$ denotes the state vector, $\textbf{y} \in \mathbb{R}^p$ denotes the measurement vector and $A \in \mathbb{R}^{n \times n}$ and $C \in \mathbb{R}^{p \times n}$ denote the state transition and measurement matrices, respectively. The time index is denoted by $k$ and it can take the values $k = 0,1,\ldots,K$, where $k = 0$ represents the initial condition and $K$ denotes the horizon. 

A discrete time LTI system such as the one described in (\ref{eq:state-space-system}) is said to be observable if the initial state $\textbf{x}_0$ can be uniquely recovered from the sequence of measurements (or sensors)\footnote{Measurements and sensors are used interchangeably in this paper.} $\left\{\textbf{y}_k\right\}_{k=0}^{K}$, where $K \geq n-1$ \cite{ybs-purple-book}. 

\subsection{Observability matrix}
\label{sec:obs-mat}
From (\ref{eq:state-space-system}), the measurement at any time index $k$ can be expressed in terms of the initial state as follows
\begin{equation}
	\textbf{y}_k = CA^{k}\textbf{x}_0 \label{eq:meas-initial-state}
\end{equation}
For a finite horizon of $K \geq n-1$ measurements, (\ref{eq:meas-initial-state}) can be expressed in vector and matrix forms as 
\begin{eqnarray}
	\left[\begin{array}{c}  \textbf{y}_0 \\ \textbf{y}_1 \\ \vdots \\ \textbf{y}_{K} \end{array}\right] &=& \left[\begin{array}{c}  C \\ CA \\ \vdots \\ CA^{K} \end{array}\right]\textbf{x}_0 \nonumber \\
	\mathcal{Y}_K &=& \mathcal{O}_K \textbf{x}_0 \label{eq:obs-lin-system}
\end{eqnarray}
forming a linear system where $\textbf{x}_0$ is the unknown (initial) state to be estimated. From linear system theory \cite{kailath80}, it is well known that $\textbf{x}_0$ has a unique solution, if and only if the matrix $\mathcal{O}_K$ -- known as the \emph{observability matrix} -- has full column rank (i.e., $\rho\{\mathcal{O}_K\} = n$).

\subsection{Observability Gramian}
\label{sec:obs-gram}
An alternative approach to deduce observability is via the observability Gramian matrix. Substituting for $\textbf{y}_k$ from the measurement equation given in (\ref{eq:meas-initial-state}), the energy of the output (measurement) over a horizon $K$ is given by
\begin{eqnarray}
	\mathcal{E}_y = \sum_{k=0}^{K}||\textbf{y}_k||^2 &=& \sum_{k=0}^{K}\textbf{x}_0^T\left(A^T\right)^kC^TCA^k\textbf{x}_0 \nonumber \\
				&=& \textbf{x}_0^T \left( \sum_{k=0}^{K}\left(A^T\right)^kC^TCA^k \right) \textbf{x}_0 \nonumber \\
				&=& \textbf{x}_0^T \mathcal{W_{O}} \textbf{x}_0 \label{eq:obs-gram-derivation}
\end{eqnarray}
Thus,
\begin{equation}
	\mathcal{W_{O}} =  \sum_{k=0}^{K}\left(A^T\right)^kC^TCA^k \label{eq:obs-gram-def}
\end{equation}
where $\mathcal{W_{O}} \in \mathbb{R}^{n \times n}$ is known as the \emph{observability Gramian} matrix. As the right hand side of (\ref{eq:obs-gram-derivation}) forms a quadratic form, it can be inferred that, for all directions of the initial state $\textbf{x}_0$ to produce an \emph{observable} output, the energy of the output should be positive, i.e., $\mathcal{E}_y > 0$. This implies that for the system to be observable, the observability Gramian matrix must be positive definite. Furthermore, using the alternate matrix form defined in (\ref{eq:obs-lin-system}) and applying it to the energy equation in (\ref{eq:obs-gram-derivation}), it is straight forward to derive the relationship between the observability Gramian and observability matrices as
\begin{equation}
	\mathcal{W_{O}} = \mathcal{O}_K^T\mathcal{O}_K \label{eq:obs-gram-mat-relation}
\end{equation}
Hence, for the LTI system in (\ref{eq:state-space-system}) to be observable, the observability Gramian $\mathcal{W_{O}}$ should have full row and column rank $n$, while the observability matrix $\mathcal{O}_K$ should have full column rank $n$. 

\subsection{Additive property of the observability Gramian}
\label{sec:additive-property}
A property that is useful in the ensuing sections is the additive property of the observability Gramian. The measurement matrix $C$ in (\ref{eq:state-space-system}) can be expressed in terms of individual measurements (or sensors) as follows
\begin{equation}
	C = \left[\begin{array}{c} C_1 \\ C_2 \\ \vdots \\ C_p \end{array} \right] \label{eq:meas-mat-breakdown}
\end{equation}
where $C_i \in \mathbb{R}^{1\times n}$ represents the $i^{th}$ row of $C$ and $i = \left\{1,\ldots,p\right\}$. Substituting from (\ref{eq:meas-mat-breakdown}) and using the expression $C^TC = \sum_{i=1}^pC_i^TC_i$, in the expression of the observability Gramian from (\ref{eq:obs-gram-def}), it is straightforward to derive the following relationship
\begin{eqnarray}
	\mathcal{W_{O}} &=& \sum_{k=0}^{K}\left(A^T\right)^kC^TCA^k = \sum_{i=1}^p \mathcal{W_{O}}^i \label{eq:obs-mat-decomp}
\end{eqnarray}
where 
\begin{equation}
	\mathcal{W_{O}}^i = \sum_{k=0}^{K}\left(A^T\right)^kC_i^TC_i A^k \label{eq:obs-gram-indiv-sensors}
\end{equation}
represents the observability Gramian of the LTI sub-system $\left(A,C_i\right)$, where $C_i$ is the row of the measurement matrix $C$ corresponding to measurement (or sensor) $i$. Hence, from (\ref{eq:obs-gram-indiv-sensors}) it can be seen that the observability Gramian of the full LTI system is additive with respect to the observability Gramian of the sub-systems derived from individual measurements.

\section{Degree of observability}
\label{sec:deg-observability}
Several candidates for observability degree based on functions of the observability Gramian are presented in \cite{krener09}, \cite{summers2015}. Observability degree measures can be broadly divided into three categories \cite{summers2015}: (a) modular functions, i.e., linear functions that map subsets or coalitions of sensors to the observability Gramian matrix, (b) submodular functions, i.e., where the contribution that each sensor makes to an existing set of sensors diminishes with increasing cardinality of the sensor set and (c) non-submodular function, where no such diminishing contribution property exists.\footnote{It is shown in \cite{summers2015} that efficient approximation strategies exist for submodular functions that significantly reduce computational complexity for sensor selection or placement posed as maximization problems.} Table \ref{tab:obs-degree-metrics} lists one commonly used observability degree metric from each of the three function types mentioned.

\begin{table}[h]
\centering
\begin{tabular}{|c|c|c|c|}
\hline
\textbf{Metric} & \textbf{Expression} & \textbf{Type} \\
\hline \hline
Trace & $tr\left\{\mathcal{W}_{\mathcal{O}}\right\}$ & Modular \\
\hline \hline
Log determinant & $\log \det \left\{\mathcal{W}_{\mathcal{O}}\right\}$ & Submodular \\
\hline \hline
Min. eigenvalue & $\lambda_{min}\left\{\mathcal{W_{O}}\right\}$ & Non-submodular \\
\hline
\end{tabular}
\caption{Commonly used measures on the observability Gramian matrix, one each from the three modularity categories discussed in \cite{summers2015}.}
\label{tab:obs-degree-metrics}
\end{table}

The trace of the observability Gramian measures the total energy captured by the finite sequence of outputs, as shown in (\ref{eq:obs-gram-derivation}). As the output energy is shown to take a quadratic form, geometrically, the observability Gramian forms an ellipse. From (\ref{eq:obs-gram-derivation}), it can be seen that larger the trace, larger the energy of the output, implying a higher degree of observability. However, there can be cases where a measurement (sensor) or a coalition of sensors lead to ``strong'' observability of certain directions while leaving other directions as unobservable.  Such cases (also known as partially observable systems) can result in a large trace even while having one or more zero eigenvalues. The minimum eigenvalue of the observability Gramian, however, is a direct indicator of the weakest observable direction \cite{ham83}, with zero indicating unobservability. It is also a much better indicator of observability degree as farther the minimum eigenvalue is away from zero, the ``more observable'' the system is. Hence, for most applications as well as from a numerical robustness aspect, it is more desirable to maximize the minimum eigenvalue of the observability Gramian than its trace during sensor selection for filter design. 

\emph{Remark 1:} For the purpose of the present paper $\log \det \left\{\mathcal{W}_{\mathcal{O}}\right\}$ is not considered, as it is undefined for sensor coalitions that result in the observability Gramian matrix becoming singular (loss of observability).

\section{Shapley values -- An overview}
\label{sec:shapley-values}
In cooperative game theory, when a game is played by a group (or coalition) of players and there is a well defined payout from the game, Shapley values \cite{shapley53} are the fair way to allocate payout contributions to individual players. Shapley value of a player $i$ is the expected contribution the player makes to the payout received \emph{by the full coalition of players}, marginalized over all the coalitions that the player can join in. In other words, Shapley values are a way to quantify the impact of a player in a multi-player game based on the value they bring to each coalition of players they can join in. The contribution of this paper is to leverage Shapley values to \emph{fairly and exactly} quantify individual sensors' contributions towards overall observability degree of an LTI system. Due to the axioms that underpin Shapley values, the contributions thus computed can be thought of as the ``fair'' allocation of importance towards observability degree for each sensor. In the rest of the section we provide an overview of the building blocks of Shapley value theory.

Suppose there is a set of players $N = \left\{1,2,\ldots,p\right\}$, the Shapley value for a player $i \in N$ is defined via a value function $v(\cdot)$ that characterizes the payout of the game. It is defined as
\begin{equation}
	\phi_i(v) = \sum_{S \subseteq N\backslash\{i\}} \frac{|S|!\left(p-|S|-1\right)!}{p!} \left[v\left(S \cup \{i\}\right) - v\left(S\right)\right] \label{eq:shapley-definition}
\end{equation}
where $\phi_i$ denotes the Shapley value of player $i$, $S$ denotes a coalition of players that is a subset of the superset $N$ and $S \backslash \{i\}$ represents a coalition $S$ that does not include player $i$. The value function $v\left(S\right)$ represents the payout attributable to coalition $S$.\footnote{The payoff from the game if only players in the subset or coalition of players $S$ were taking part.} Marginal contribution of a player $i$ to a coalition $S$ is given by $v\left(S \cup \{i\}\right) - v\left(S\right)$, i.e., the value added by the player $i$ to the coalition $S$.  Hence, the Shapley value of player $i$ is nothing but the average change in payout that $i$ causes by joining all possible coalitions, as compared to the payouts when it is not part of them. The weight term in (\ref{eq:shapley-definition}) is the fraction of all possible ways in which the players in $S$ can precede $i$.

A Shapley value is the only attribution method (or the \emph{fair} attribution method) that satisfies the following four axioms \cite{shapley53}:
\begin{itemize}
	\item[1.] \emph{Efficiency}: The total value of the game is fully distributed, i.e., the Shapley values of the individual players should add up to the overall payout of the game:
			\begin{equation}
				\sum_{i \in N} \phi_i(v) = v(N) 
			\end{equation}
			where $N = \left\{1,2, \ldots, p\right\}$ is the set of all players. This makes Shapley values very interpretable -- each Shapley value is simply the decomposition of the total payout to individual players.
	\item[2.] \emph{Symmetry}: Players who contribute equally, get equal value. If,
			\begin{equation}
				v\left(S \cup \{j\}\right) = v\left(S \cup \{k\}\right) 
			\end{equation}
			$\forall S \subseteq N\backslash\{j,k\}$, then $\phi_j(v) = \phi_k(v)$. 
	\item[3.] \emph{Dummy}: A player who adds no value to any coalition gets zero. If,
			\begin{equation}
				v\left(S \cup \{j\}\right) = v\left(S\right) 
			\end{equation}
			$\forall S \subseteq N\backslash\{j\}$, then $\phi_j(v) = 0$. The interpretation is that a player that does not add value to \emph{any} coalition contributes nothing to the total payout of the game. Such players are called dummy players.
	\item[4.] \emph{Additivity}: If a game with payout or value $v^*$ can be expressed as a sum of two individual games with payouts $v_1$ and $v_2$, i.e., $v^* = v_1+v_2$, then the Shapley value of the player $j$ can also be expressed as 
			\begin{eqnarray}
				\phi_j(v_1 + v_2) = \phi_j(v_1) + \phi_j(v_2) \nonumber \\
				\phi_j(v^*) = \phi_j(v_1) + \phi_j(v_2) 
			\end{eqnarray}
		as the sum of the Shapley values of the player in those individual games. If the value function $v^*$ can be shown to be additive over the value functions of individual models (for e.g., ensemble approaches such as bagging), then it can be verified that the additivity axiom is met \cite{molnar2025}.
\end{itemize}

The present paper proposes the use of Shapley values to \emph{fairly} allocate contributions to individual sensors (players) towards the degree of observability (overall payout) based on certain value functions (discussed in Section \ref{sec:deg-observability}) that characterize the degree of observability of LTI systems.

\section{Individual sensor contribution towards degree of observability}
\label{sec:shapley-observability}

In this section, we show how the two of the observability degree metrics presented in Table \ref{tab:obs-degree-metrics}, i.e., the trace and the minimum eigenvalue of the observability Gramian, can both be used as value functions to derive Shapley values. We omit the log determinant of the observability Gramian from the scope of the present paper as, even though it is a submodular function with known approximation properties, it is undefined for sensor coalitions that result in loss of rank. 

We now show how the chosen observability degree measures can be ``broken down'' into allocations that provide a fair quantification of the contribution of individual sensors -- characterised by rows $C_i$ of the measurement matrix as shown in (\ref{eq:meas-mat-breakdown}) -- via Shapley values.

\subsection{Trace of the observability Gramian}
\label{sec:trace-metric}
As the trace of a matrix is a linear operator and using the additive property of the observability Gramian in (\ref{eq:obs-mat-decomp}), it can be seen that the trace of the observability Gramian of the overall system $(A,C)$ is the sum of the traces of the observability Gramian of sub-systems formed by individual sensors $\left(A,C_i\right):$
\begin{eqnarray}
	tr\left\{ \mathcal{W_{O}} \right\} &=& tr\left\{ \sum_{i=1}^p \mathcal{W_{O}}^i\right\} \nonumber \\
							    &=&  \sum_{i=1}^p tr\left\{ \mathcal{W_{O}}^i\right\} \label{eq:trace-addivity}
\end{eqnarray}
Due to the additivity property of the trace, as shown in (\ref{eq:trace-addivity}), and since $tr\{A\} \in \mathbb{R}^1$ when $A \in \mathbb{R}^{n \times n}$ is a well defined and smooth value function, Shapley values derived by substituting the following value function in (\ref{eq:shapley-definition})
\begin{equation}
	v(S) = tr\left\{ \mathcal{W_{O}}^S \right\} \label{eq:trace-value-fun}
\end{equation}
follow all the axioms listed in Section \ref{sec:shapley-values}, as applicable. The term $\mathcal{W_{O}}^S$ denotes the observability Gramian formed by the system $\left(A,C_S\right)$ where $C_S$ is the measurement matrix formed by the coalition of sensors $S \subseteq \left\{1,2,\ldots,p\right\}$. 

Incidentally, due to its modular nature of this observability degree metric \cite{summers2015}, the following Proposition results:
\newline
\emph{Proposition 1:} \label{prop-shapley-trace-additivity}
When the value function over the full sensor set $N$ is additive over value functions over individual sensors $j$, i.e.,
\begin{equation}
	v(N) = \sum_{j=1}^pv(\{j\})	\label{eq:trace-additive-axiom}
\end{equation}
then the Shapley value of the sensor $j$ is the same as the value function over sensor $j$, i.e.,
\begin{equation}
	\phi_j(v) = v\left(\{j\}\right)	\label{eq:shapley-trace-equality}
\end{equation}

\emph{Proof:} Applying the definition of the Shapley value of sensor $j$ from (\ref{eq:shapley-definition}), we get
\begin{equation}
	\phi_j(v) = \sum_{S \subseteq N\backslash\{j\}}w_S \left[v\left(S \cup \{j\}\right) - v\left(S\right)\right] \label{eq:shapley-definition1}
\end{equation}
where $w_S = \frac{|S|!\left(p-|S|-1\right)!}{p!}$ is the weight attached to coalition $S$. From (\ref{eq:trace-additive-axiom}), it follows that
\begin{equation}
	v\left(S \cup \{j\}\right) = v\left(S\right) + v(\{j\}) \label{eq:value-differential}
\end{equation}
Substituting from (\ref{eq:value-differential}) in (\ref{eq:shapley-definition1}), we get
\begin{equation}
	\phi_j(v) = \sum_{S \subseteq N\backslash\{j\}}w_S v(\{j\}) \label{eq:shapley-simplify1}
\end{equation}
Further simplifying
\begin{eqnarray}
	\phi_j(v) &=& v(\{j\})\sum_{S \subseteq N\backslash\{j\}}w_S \nonumber \\
		      &=& v(\{j\}) \label{eq:shapley-simplify2}
\end{eqnarray}
as $\sum_{S \subseteq N\backslash\{j\}}w_S =1$. This is because sensor $j$ can join all possible coalitions $S$ that it is not a part of, without any limitations. Furthermore, as a corollary from (\ref{eq:shapley-simplify2}), the \emph{Efficiency} axiom can also be verified as follows:
\begin{eqnarray}
	\sum_{j=1}^p \phi_j(v) &=& \sum_{j=1}^pv(\{j\}) \nonumber \\
					  &=& v(N) \label{eq:verify-efficiency-axiom}
\end{eqnarray}
due to the additivity property in (\ref{eq:trace-additive-axiom}). Proposition 1 thus implies that the Shapley values derived from the trace of the observability Gramian, \emph{do not take into account interactions between individual sensors}. We now investigate if value functions based on minimum eigenvalue of the observability Gramian, yields Shapley values that take into account interaction effects where, even though a single sensor by itself might not make the system observable when part of coalitions with other sensors it can both make the system observable and improve the degree of observability.

\subsection{Minimum eigenvalue of the observability Gramian}
\label{sec:eigenvalue-metric}
As discussed in Section \ref{sec:deg-observability}, for the observability Gramian to be positive definite, its minimum eigenvalue should be positive, i.e., $\lambda_{min}\left\{\mathcal{W_{O}}\right\} > 0$. From (\ref{eq:obs-mat-decomp}), we get
\begin{equation}
	\lambda_{min}\left\{ \mathcal{W_{O}} \right\} = \lambda_{min}\left\{ \sum_{i=1}^p \mathcal{W_{O}}^i\right\} \label{eq:eigenvalue-addivity1}
\end{equation}
However, in general, due to the non-additive nature of eigenvalues
\begin{equation}
	\lambda_{min}\left\{ \sum_{i=1}^p \mathcal{W_{O}}^i\right\} \neq  \sum_{i=1}^p \lambda_{min}\left\{ \mathcal{W_{O}}^i\right\} \label{eq:eigenvalue-addivity}
\end{equation}
As a result, contrary to Proposition 1, we cannot prove that the Shapley values of the individual sensors are identical to the value functions over individual sensors. Hence, they \emph{do} consider interaction effects with other sensors and \emph{reward situations where a sensor increases the minimum eigenvalue of the resulting observability Gramian by joining one or more coalitions}. Furthermore, as is shown in \cite{summers2015}, since the minimum eigenvalue of the observability Gramian is a non-submodular function, it does not follow the property of diminishing returns. This implies that there could be scenarios where marginal contribution of a given sensor could be higher for a larger coalition of sensors than a smaller one. As a result, this metric is able to identify directions in the state space whose unobservable mode(s) become observable due to the addition of a given sensor, better. Also, as $\lambda_{min}\left\{ \mathcal{W_{O}}\right\}$ is a valid value function that maps a matrix in $\mathbb{R}^{n\times n}$ to a scalar in $\mathbb{R}^{1}$ due to its symmetricity and because it is real, finite and well defined for all sensor coalitions considered, Shapley values can be derived and the axioms listed in Section \ref{sec:shapley-values} hold where applicable. 

\section{Simulations and results}
\label{sec:sim-results}

Scenario I considers the case where no individual sensor (or measurement), by itself, results in full observability, but the overall system is still observable. Scenario II considers an example where some sensors, by themselves, yield full observability while some do not.

\subsection{Scenario I: Collaborative sensors}
\label{sec:collab-sensors}
Consider the following two-state, two-measurement system
\begin{eqnarray}
	A = \left[\begin{array}{cc} 1 & 0 \\ 0 & 1 \end{array} \right] &;& 
	C =  \left[\begin{array}{cc} 1 & 1 \\ 1 & -1 \end{array} \right] \label{eq:obs-example}
\end{eqnarray}
where both the ``sensors'' $C_1$ and $C_2$ (the two rows of the $C$ matrix) observe a linear combination of the two states. It is straightforward to verify that this system is fully observable even though, if the two sensors ($C_1$ and $C_2$) are used individually, the observability Gramian suffers a rank loss. This is a classic case of both sensors being equally important, as individually even though they may not have a high degree of observability, but together they make the system observable.\footnote{This class of sensors are also known as complementary sensors.}
\begin{table}[h]
\centering
\begin{tabular}{|c|c|c|c|}
\hline
\textbf{Sensor} & \textbf{Value Function} & \textbf{Standalone Value} & \textbf{Shapley Value} \\
\hline
$C_1$ & Trace & 20.0 & 20.0 \\
$C_2$ & Trace & 20.0 & 20.0 \\
\hline \hline
$C_1$ & Min Eigenvalue & 0 & 10.0 \\
$C_2$ & Min Eigenvalue & 0 & 10.0 \\
\hline
\end{tabular}
\caption{Standalone and Shapley values for two sensors using two different observability metrics with horizon $N=10$.}
\label{tab:scenario1}
\end{table}

From Table \ref{tab:scenario1}, we can see that; (a) as is expected from Proposition 1, the standalone sensor traces and the Shapley values are identical and (b) the minimum eigenvalues if the two sensors were used individually result in a loss of observability, but their Shapley values are identical\footnote{\emph{Symmetry} axiom of Shapley values -- due to equal importance.} and show their fair importances towards the eigenvalue based observability degree metric. This shows that the minimum eigenvalue metric captures the interaction effect between the two sensors whereas the trace of the observability Gramian does not. It can also be verified that the Shapley values of the two metrics sum up to the overall observability degrees, respectively, illustrating the effect of the \emph{Efficiency} axiom.

\subsection{Scenario II: Standalone, collaborative and redundant sensors}
Consider a slightly more complex example in the form of the following three state, four sensor system
\begin{eqnarray}
	A = \left[\begin{array}{ccc} 1 & 1 & 0 \\
						0 & 1 & 1 \\
						0 & 0 & 1 \end{array} \right] &;& 
	C =  \left[\begin{array}{ccc} 1 & 0 & 0 \\  
						0 & 1 & 0 \\  
						1 & 1 & 0 \\  
						0 & 0 & 1 \end{array} \right] \label{eq:obs-example-scen2}
\end{eqnarray}
From visual inspection of (\ref{eq:obs-example-scen2}), it can be inferred that sensor $1$ alone, i.e., $C_1 = \left[\begin{array}{ccc} 1 & 0 & 0\end{array} \right]$, is enough to get full observability. Furthermore, individually used, sensors $C_2 = \left[\begin{array}{ccc} 0 & 1 & 0\end{array} \right]$ and $C_4 = \left[\begin{array}{ccc} 0 & 0 & 1\end{array} \right]$ would lead to a rank loss. Sensor $3$, i.e., $C_3 = \left[\begin{array}{ccc} 1 & 1 & 0\end{array} \right]$ can be thought of as a redundant sensor to the $C_1, C_2$ pair as it observes a linear combination of the first two states. Intuitively, it can be said that as $C_1$ leads to the observability of the system by itself, it should have a high Shapley value. Also, $C_3$ is a redundant sensor that can be used in combination with $C_1$ to eliminate the need for both $C_2$ and $C_3$.  Sensor $C_4$, though, adds the least value to any well constructed observability index as the first three sensors are enough to observe the third state. 
\begin{table}[h]
\centering
\begin{tabular}{|c|c|c|c|}
\hline
\textbf{Sensor} & \textbf{Value Function} & \textbf{Standalone Value} & \textbf{Shapley Value} \\
\hline
$C_1$ & Trace & 3187.0 & 3187.0 \\
$C_2$ & Trace & 295.0 & 295.0 \\
$C_3$ & Trace & 5312.0 & 5312.0 \\
$C_4$ & Trace & 10.0 & 10.0 \\
\hline \hline
$C_1$ & Min Eigenvalue & 1.3920 & 1.5129 \\
$C_2$ & Min Eigenvalue & 0.0000 & 0.2306 \\
$C_3$ & Min Eigenvalue & 0.6405 & 0.7089 \\
$C_4$ & Min Eigenvalue & 0.0000 & 0.0243 \\
\hline
\end{tabular}
\caption{Standalone and Shapley values for four sensors using trace and minimum eigenvalue of observability Gramian with horizon $N=10$.}
\label{tab:scenario2}
\end{table}

From Table \ref{tab:scenario2}, it can be verified that, as expected from Proposition 1, the Shapley values based on trace as the value function do not take into account interaction effects. If trace is used as the value function, the Shapley value of $C_3$ turns out to be the highest which is not intuitive as discussed earlier. Shapley values based on the minimum eigenvalue value function, however, achieve a very intuitive allocation of sensor importances. Firstly, the Shapley value of each sensor shows a slight increase over the standalone value functions. This is due to the interaction effects that they capture which the trace based value function does not. Secondly, aligning with intuition, $C_1$ has the highest Shapley value followed by $C_3$. Note that the orders are reversed when trace is used as the value function. The Shapley values of $C_2$ and $C_4$ also show a small increase, as even if (by themselves) they do not make the system observable but when joined in coalitions with other sensors, they contribute towards improving the degree of observability. Due to the \emph{Efficiency} axiom, the sum of the Shapley values based on the minimum eigenvalue can be verified to be the minimum eigenvalue of the overall system, i.e., 2.477, with Shapley values of sensors $C_1$ and $C_3$ alone responsible for 90\% of the observability degree even though they make up just 50\% of the full sensor set.

\section{CONCLUSIONS}
This paper presents an approach for fair allocation of importance of individual sensors towards overall degree of observability. This has applications in sensor selection/ placement problems and filter design for state estimation. It is shown that fair allocation of contributions from each sensor in an LTI system towards degree of observability can be obtained by deriving Shapley values. This is illustrated using two commonly used observability degree metrics, (a) trace of the observability Gramian and (b) minimum eigenvalue of the observability Gramian. The relative merits of using either of the two metrics are highlighted via connection to the Shapley value axioms. It is shown that minimum eigenvalue of the Gramian is more reliable for both overall observability degree measurement as well as sensor contribution allocation via Shapley values as it measures interaction effects between sensors.

\addtolength{\textheight}{-12cm}   








\end{document}